# The Predictive Power of Chemical Bonding Analysis in Materials: a Perspective on Optoelectronic Properties

Gabriele Saleh[a,*], Liberato Manna[a,*]

[a] Nanochemistry Department, Istituto Italiano di Tecnologia, via Morego 30, 16163 Genova, Italy.

**ABSTRACT:** Chemical bonding governs how atoms interact to form compounds, thereby determining their physicochemical properties. Despite being an elusive concept, chemical bonding has led to the development of models and tools to explain and predict the behavior of chemical species. This perspective addresses the adoption of chemical bonding analysis to the study of optoelectronic materials, emphasizing the importance of its predictive aspect. After reviewing the evolution of chemical bonding models from the first Lewis formulation to the present day, the perspective discusses material classes and chemical bonding phenomena most relevant for light harvesting and emission. We delve into metal halide perovskites and structurally related materials, given their central role in optoelectronic research. Various aspects of chemical bonding in these materials are surveyed, from the structure-property relationship to the rationalization of their electronic properties through molecular orbital diagrams. Two chemical bonding features are particularly important for optoelectronic materials: the $ns^2$ lone pairs of the cations typically found in these materials (e.g. Pb, Sb, Bi) and the antibonding nature of valence and/or conduction bands. We discuss in depth the models to predict the implications of these two phenomena on optoelectronic properties. We also explore chalcohalides, a class of materials whose optoelectronic properties are recently emerging. From the chemical bonding perspective, these materials display intriguing phenomena due to the interplay of various types of chemical bonds. Finally, we discuss our vision on the role of chemical bonding analysis in the future of materials science, including synergies and antitheses with machine learning.

**1. Introduction**.

Chemical bonding is arguably a central paradigm in modern chemistry. In both fundamental and applied chemistry, nearly every question—reaction mechanisms, compound stability, crystal and molecular structures, etc.—is framed in terms of the formation or breaking of chemical bonds. Despite its central importance, the idea of chemical bonds eludes any rigorous definition[1,2]. And yet, the whole body of *understanding* the chemistry community has about substances makes use of concepts related to chemical bonding[3–5]. Importantly, the application of these concepts allows a chemist to make qualitative *predictions* on the properties of chemical species (molecules and solids). The simplest form of this approach is to directly apply heuristic notions of chemical bonding, for example gauging the dipole of a molecule based on the electronegativity of its constituting atoms[6] or ranking the melting temperature of a set of solids based on bond strengths considerations[7]. With the advent of computational chemistry and advanced spectroscopic techniques, chemical bonding concepts were integrated with and underpinned by the electronic structure of chemical species, enabling more sophisticated predictions. An early renowned example is the prediction of the outcome of pericyclic reactions based on orbital symmetry considerations and simple electronic structure calculations[8]. All these approaches are generally referred to as "chemical bonding analysis". Broadly speaking, this analysis consists in the application of those concepts and models that allow chemists to describe how atoms and molecules interact together: bonding/antibonding states, ionic/covalent bond character, lone pairs, atomic radii, to name just a few from many.

This perspective is concerned with the adoption of chemical bonding analysis to understand and predict the physicochemical properties of materials. In particular, the discussion focuses on inorganic materials for optoelectronic applications, typically light harvesting, detection, and emission. Our choice of this class of materials is dictated by two reasons. First, their relevance to the scientific community. In fact, these materials are at the heart of important technologies needed to tackle the sustainability challenge, such as solar panels and low-footprint light sources. The second reason is their complexity, which represents a challenging test bench for chemical bonding analysis. Indeed, the performance of optoelectronic materials hinges on a specific set of concurrent properties such as absorption coefficient, carrier mobility and lifetime, and thermal conductivity[9–11]. At the atomic level, these properties depend, often in a non-trivial manner, on the electronic structure (ground and excited states) of the material and on its vibrational dynamics. These electronic and vibrational factors are determined by the specific way in which atoms interact, *i.e.* by the chemical bonding pattern of the material (Fig. 1).

In this perspective, we make the case that chemical bonding analysis, beside rationalizing the properties of materials, can (and should) be adopted as a predictive tool. To this aim, we start by concisely surveying the historical development of chemical bonding analysis, emphasizing the role of predictive power (Sect. 2). This is complemented by a discussion of the chemical bonding analysis in solids, presenting the tools that are mostly used in the studies discussed in this perspective (Sect. 3). In the main body of this work (Sects. 4-7), material types and chemical bonding phenomena that are most relevant for optoelectronic applications are discussed. Concerning material classes, we focus on the well-established metal halide perovskites (Sect. 4), and the recently emerging chalcohalides (Sect. 7). In these and related materials, it turns out that two chemical bonding features are particularly relevant for optoelectronic performance. These are the $ns^2$ lone pairs of cations and the antibonding nature of the valence and conduction bands, which are discussed in Sects. 5 and 6, respectively. For these four sections, we selected those studies that go beyond the simple description of the electronic structure of materials. That is, we focus on those works that analyze chemical bonding with the aim of guiding materials design, either by deriving rules to predict the properties of materials that are relevant to their technological performance, or by devising screening criteria to identify new materials for a given application. Finally, Sect. 8 discusses possible research directions and contextualizes the presented analysis in the modern era of data-driven science.

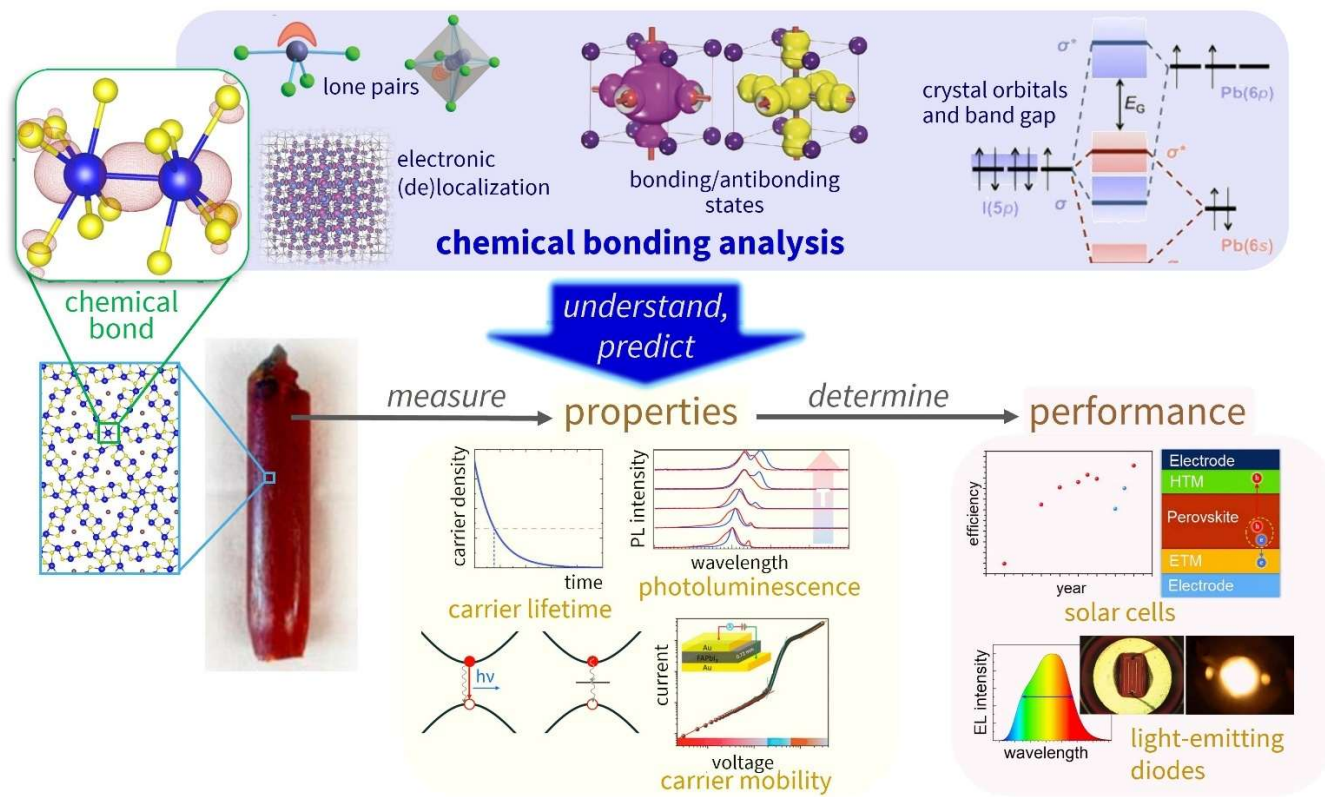

**Figure 1.** Conceptual illustration of how chemical bonding analysis can foster the design of materials for optoelectronic applications by allowing chemists to understand and predict their properties. Images adapted from ref [12] (crystals), refs [13–17] (chemical bonding panels), refs. [18] (properties), and refs. [19,20] (performance). Images of refs [12,16] reproduced with permission of Springer Nature. Image from refs. [17,18] used with permission of John Wiley & Sons. Images of ref [19] used with permission of the Royal Society of Chemistry. Permissions conveyed through Copyright Clearance Center, Inc..

## 2. The predictive power of chemical bonding analysis: a historical perspective.

The idea of bonds joining atoms started taking shape during the 19$^{th}$ century[21], setting the stage for the formal chemical bonding theory of G. N. Lewis, published in 1916[22] and still taught today. Remarkably, this happened before the discovery of quantum mechanics. It was the 1927 seminal work of Heitler and London[23] that cast the concept of chemical bonding into its quantum mechanical origin. Afterwards, scientists such as Mulliken, Hund, and most prominently Pauling, merged the chemical, more empirical and (quantum) physical descriptions of the chemical bonding into an interdisciplinary subject[24,25]. The language of chemists was then enriched with many new concepts such as electronegativity, resonance, and electron delocalization, which were further refined and expanded as the electronic structure of compounds became accessible through computer simulations. All these theories and models, whose evolution has been summarized here in a nutshell (see, e.g. ref. [24] for a more complete account), were developed with the aim of rationalizing and predicting the chemical behavior of elements and compounds. In fact, even their rigorous connection to reality was secondary. For example, Lewis'[22] (and Langmuir's[26]) cubic atom model described the valence electrons of an atom as static particles positioned on the vertices of a cube. Even at that time the model was not considered realistic[25], but it could explain the composition and reactivity of chemical species, and that was what mattered to the chemical community. Charles Coulson, the author of one of the most influential books on bonding[4], even labelled chemical bonds as "a figment of our own imagination"[27]. In fact, as highlighted by Frenking *et al.*[28], it is fundamental to distinguish between the *mechanism* of bonding and the *models* of chemical bonding. The former is a strictly quantum mechanical phenomenon[28,29], whose understanding alone does not generally entail predictive power. The models are instead built to be predictive, at least in principle. As the famous aphorism goes, *all models are wrong, but some are useful*[30].

During the past few decades, a profusion of mathematical tools has been developed to analyze chemical bonding based on either molecular orbitals or electron density (QTAIM[31], ELF[32], EDA[33], NBO[34], DAFH[35], etc.). The aim of these approaches is to extract, from the complexity of the electronic structure, a description of the studied system in terms of the concepts that form the language of chemists, such as bonds, lone pairs, and atomic charges. However, while this endeavor brought important physical insights into the mechanism of chemical bonding, much of the original pragmatism of theoretical chemistry was lost. That is, the strong orientation towards explaining the observed phenomena gave way to more fundamental discussions on the nature and quantum mechanical mechanisms of chemical bonding. Quintessential examples are the "3c-2e" bond in diborane, first proposed in 1945[36] and still under discussion[37,38], and the ionic *vs* covalent nature of the Li-F bond[39,40]. These sorts of fundamental discussions led to numerous debates (see *e.g.* [41–46]), sometimes fierce[47,48]. Applications of chemical bonding analysis tools to the explanation and prediction of experimental properties are comparatively much rarer, even more so for solid state chemistry.

This section provided an overview of how chemical bonding models evolved from the early stages to the modern era. We maintain that, in present-day materials science, chemical bonding analysis is largely adopted as a descriptive tool, while its predictive aspect remains much less developed. Exceptions to this trend clearly exist, as those discussed in this work. In general, we argue that the time is ripe, in terms of both computational power and data availability, to systematically develop the predictive aspect of chemical bonding analysis (see also Sect. 8), lest it becomes a form of academic solipsism.

### 3. Chemical bonding in optoelectronic materials: methods and tools.

In solid state, the language of chemical bonding differs to some extent from that originally developed for molecules. To account for the unit cell periodicity, the electronic states are typically represented through band diagrams and density of states (DOS)[49]. Nonetheless, the chemical bonding concepts developed for molecules can be extended to solids by complementing them with information about periodicity, as eloquently explained by R. Hoffmann[50]. Interestingly, most of the works discussed in the next sections are based on the identification of electronic states as bonding, antibonding, or nonbonding (lone pairs). In some cases, (periodic) molecular orbital (MO) diagrams are built, exploiting the fact that the materials discussed in this perspective are semiconductors and as such they retain a certain degree of discrete energy levels also in solid state (unlike metals and alloys). For the above-mentioned identification of electronic states, beside plotting the contribution of individual atomic orbitals to the DOS (partial DOS or p-DOS), the following computational tools are typically adopted:

- *Partial electron density plots in real space*. The electron density associated to some particular band or -more often- to a given energy range of the DOS is plotted. By visual inspection, it is easy to identify electronic states as bonding (accumulation of electron density between atoms), antibonding (nodal plane in between atoms), or lone pairs (electron density concentrated close to the atom). This procedure, especially if coupled with p-DOS analysis, makes it possible to recast the band structure into a solid-state MO diagram (*e.g.* Figs. 2,3, discussed in Sect. 4). Furthermore, in materials for light absorption/emission, this method conveys important information on electronic excitations: the electron density distribution of the edge states (*i.e.* those close to the band gap) represents a good approximation for the real-space distribution of the excited electron and the hole.

- *Crystal orbital plots*. A crystal orbital is the solid-state equivalent of molecular orbitals[51]. Plotting the real-space distribution of crystal orbitals provides similar information as the partial electron density, although the orbitals are more cumbersome to analyze since there can be many orbitals in a small DOS energy range, especially if the unit cell is large and/or there are flat bands.

- *Crystal orbital overlap population (COOP)*[52]. It is a solid-state extension of the well-known Mulliken overlap population[53], which measures the electron sharing between two atoms. For each pair of atomic orbitals ν and μ, COOP is expressed as:

$$COOP(\varepsilon) = S_{\mu\nu} \sum_{i,k} c^*_{\mu,i,k} c_{\nu,i,k} \delta(\varepsilon_i(k) - \varepsilon) \quad (1),$$

where $S_{\mu\nu}$ is the overlap between the atomic orbitals, $c_{\nu,i,k}$ is the coefficient of the ν-th atomic orbital at the i-th band at the reciprocal space point k; δ is a Dirac delta function that selects only bands and k points at the energy at which COOP is calculated. The result is an energy-resolved, DOS-like plot in which, for a given pair of atoms (or atomic orbitals), the electronic states are identified as bonding (positive COOP, charge accumulated between atoms) or antibonding (negative COOP, charge depleted from the interatomic region). It can be viewed as a quantitative analysis of the partial electron density analysis presented above.

- *Crystal orbital Hamilton population (COHP)*[54]. Its mathematical form is similar to COOP, but the overlap matrix is replaced by the Hamiltonian matrix $H_{\mu\nu}$:

$$COHP(\varepsilon) = H_{\mu\nu} \sum_{i,k} c^*_{\mu,i,k} c_{\nu,i,k} \delta(\varepsilon_i(k) - \varepsilon) \quad (2).$$

The introduction of the Hamiltonian matrix brings the interatomic interaction description on an energy basis, thereby making a direct connection to the bonds and material stability.

Detailed pedagogical derivations of COOP and COHP can be found in refs [51] and [55].

Being computational in nature, all these descriptors are affected by the adopted methodology, typically the DFT functional[56] and the basis set[57]. As this work focuses on (mostly qualitative) chemical bonding analysis, we do not discuss this dependence in depth, we only highlight here a few key aspects. We do not expect any of these descriptors to have a strong method dependence, except for those cases which are generally challenging for DFT, typically strongly correlated materials[58] (not treated in this work). Only the band gap magnitude requires the adoption of appropriate DFT functionals to match experimental values[59]. As for COHP, it was shown not to have

## 4. Chemical bonding in metal halide perovskites.

We now move to practical examples of chemical bonding analysis, starting with a discussion on one of the most studied classes of materials for optoelectronic applications: metal halide perovskites. They are compounds having the crystal structure shown in Fig. 2a, with general formula $ABX_3$ (X=Cl,Br,I), where A and B are typically a monovalent cation and a divalent metal, respectively. This family of materials, and in particular $CsPbX_3$ and $CH_3NH_3PbX_3$, are exploited for numerous technological applications, including solar cells, light-emitting diodes (LEDs), various types of detectors, and thermoelectrics[61–63]. At the atomic and electronic level, $ABX_3$ perovskites display some properties that are known to be pivotal for those applications, namely: solar-active and tunable band gap, defect tolerance, exceptionally long carrier lifetimes, and high photoluminescence quantum yield[64]. These properties, some of which may also be found in the structurally related "nD metal halides" (n=0-2)[65], are determined by the electronic structure of the materials. This, in turn, is a consequence of their chemical bonding pattern. Therefore, chemical bonding analysis can uncover the mechanism behind their optimal properties, thereby paving the way for the rational design of new optoelectronic materials.

The MO diagram of $APbX_3$ perovskites, first reported in 2003[67] and further investigated in several other studies (e.g. refs [16,66,68]), typically looks like Fig. 2b. The A cation, while playing a key structural role, gives no contribution to the electronic states at or near the band edges[66,69]. Both the valence band maximum (VBM) and the conduction band minimum (CBM) are Pb-X antibonding states[66]. This is one of the key features of the chemical bonding pattern of $APbX_3$ that make their electronic structure uniquely suited for optoelectronic and light-emitting applications (see also Sect. 5). Another key feature is the electronic configuration of Pb, with its stable $6s^2$ lone pair (see also Sect. 6). In fact, by comparing the band structure diagrams of a series of $CsMBr_3$ (M = Pb, Cd, Sr) materials and so-called double-perovskites $CsM'M''Br_3$ (M'/M''=Tl/Bi, Ag/Bi, Ag/In, K/Bi, K/In), Fabini et al.[70] illustrated how the electronic structure features that make $AMX_3$ ideal for applications are unique for M=Pb. In particular, none of the studied materials could match the band gap width, bands dispersion, and direct band gap character of lead halide perovskites[*].

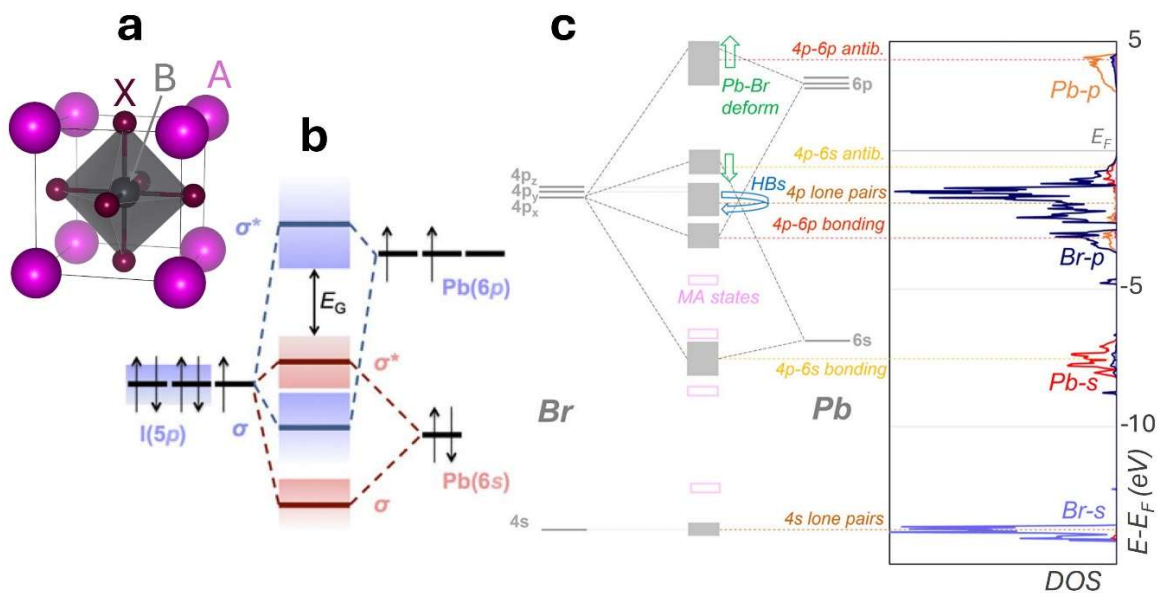

**Figure 2.** Structure and bonding of lead halide perovskites. (a) crystal structure of cubic $ABX_3$ perovskite. (b) simplified MO diagram of $ABX_3$ perovskites near band edges. (c) DOS and corresponding MO diagram of $MAPbBr_3$ (MA=$CH_3NH_3^+$). The effect of hydrogen bonds (HBs) and Pb/Br lattice deformation on the energy levels is pictorially indicated. Images adapted from refs. [16] (b; reproduced with permission from Springer Nature) and [66] (c).

Goesten and Hoffmann[71] analyzed the band structure of (cubic) $CsMX_3$ (M = Sn, Ge, Pb; X = F, Cl, Br, I) in detail. They studied how the interaction among atomic orbitals changes in the various points of reciprocal space, thereby qualitatively predicting the energy of crystal orbitals. COOP plots were adopted to support that analysis. Through this qualitative analysis they were able to rationalize the shape of all bands in a wide energy range around the Fermi level. The *understanding* thus created allowed them to predict the band gap changes induced by the variation in composition within the studied family of materials. This meticulous study shows how a detailed chemical bonding analysis can reinterpret the band structure diagram of a fairly complex material into a simple yet predictive model accessible through undergraduate-level theoretical chemistry knowledge.

Once the electronic structure of $AMX_3$ perovskites is rationalized in terms of chemical bonding concepts, the chemist gains a powerful tool to tune their properties through compositional changes or doping. A notable example of this approach was provided by Walsh[69], who discussed the chemical bonding pattern of MAPbI$_3$ (MA= CH$_3$NH$_3^+$) and adopted it as a reference material to gauge the technological implications of substituting any of the three compositing elements. Chen et al.[72] adopted chemical bonding considerations to address one of the fundamental issues of AMX$_3$: ions migration, which triggers the material degradation. They showed that Ni and Mn doping creates an energetic barrier for Br vacancy hopping and, by COHP and DOS analysis, attributed it to a "long-range lattice stabilization" induced by a transfer of electrons from the Pb(6s)-Br(4p) antibonding states to the dopant 3d orbitals. While we do not fully agree with their interpretation of the results, their chemical bonding analysis complements the energetic profiling by clearly showing that the Ni-Br bond is stronger than Pb-Br, which creates an effective barrier for Br vacancies migration.

In ABX$_3$ perovskites, the B-X scaffold undergoes local distortions, which directly affect their electronic properties. Saleh et al.[66] analyzed this effect in CH$_3$NH$_3$PbBr$_3$, aiming to explain the double peak in its photoluminescence spectra. The distortions in the Pb-Br framework, and the ensuing phase transitions, were demonstrated to be governed by the temperature-induced rotations of the methylammonium molecules (CH$_3$NH$_3^+$). Importantly, the effect of these rotations on the molecular orbital diagram, derived from the electronic structure analysis, was studied (Fig. 2c). It was shown how N-H⋯Br hydrogen bonds lower the energy of the Br lone pairs, without directly affecting the band gap. Instead, the distortion of the Pb-Br framework was demonstrated to enlarge the band gap. Curiously, they showed that this effect goes in a direction opposite to what is predicted by atomic orbitals overlap considerations. This indicates that more complex mechanisms such as the electron-electron interaction across crystal orbitals are responsible for the observed band gap changes. These insights were then extrapolated to the mesoscale and compared with available experimental data, thereby inferring that the band gap of CH$_3$NH$_3$PbBr$_3$ displays spatial and time fluctuations, which are responsible for their peculiar photoluminescence spectra.

Ganose and co-workers gave an insightful account of chemical bonding in vacancy-ordered double perovskites[73,74], a class of promising materials for optoelectronic applications[75]. These are compounds of general formula A$_2$BX$_6$, whose structure is obtained from that of perovskites by doubling the unit cell and removing every second B cation (Fig. 3a). Clearly, these structural and compositional changes with respect to perovskites reflect on the oxidation state of the B cation and, consequently, the nature of the band edge states. When B cations belong to group IV (e.g. Sn), the CBM is formed by B(s)-X(p) antibonding orbitals, while the VBM is formed by the p lone pairs of the X halogen. It is only when B cations belong to group VI (e.g. Te) that both VBM and CBM acquire the B-X antibonding character observed in ABX$_3$ perovskites (Fig. 3b), although the band structure is still different due to different stoichiometry and geometry. Based on this thorough understanding, Maughan et al.[74] could predict the changes in the band gap width and type of A$_2$BX$_6$ for various B and X elements. The change of the band gap width with the nature of the halogen atoms in Cs$_2$SnX$_6$ (I<Br<Cl) is determined by the energy difference between their p orbitals and the 5s orbitals of Sn: the smaller this difference, the more effective is the S-X hybridization. A stronger hybridization makes the energy difference between the X lone pairs (VBM) and the Sn(s)-X(p) antibonding orbitals (CBM) larger, thus increasing the band gap. The substitution of Sn with Te changes the band gap from direct to indirect, and this could be explained by considerations on atomic orbital overlap at different k points, similarly to the above mentioned study on CsBX$_3$[71]. The plot of the electron density distribution of VBM and CBM led to an important observation to understand their properties: despite BX$_6$ octahedra are structurally isolated, they interact electronically, which explains the significant dispersion of the bands. The understanding created in this study, and in particular the molecular orbital diagram(s), was exploited by Tan et al.[76] to explain the boost in the photoluminescence of Rb$_2$SnCl$_4$ upon Te doping.

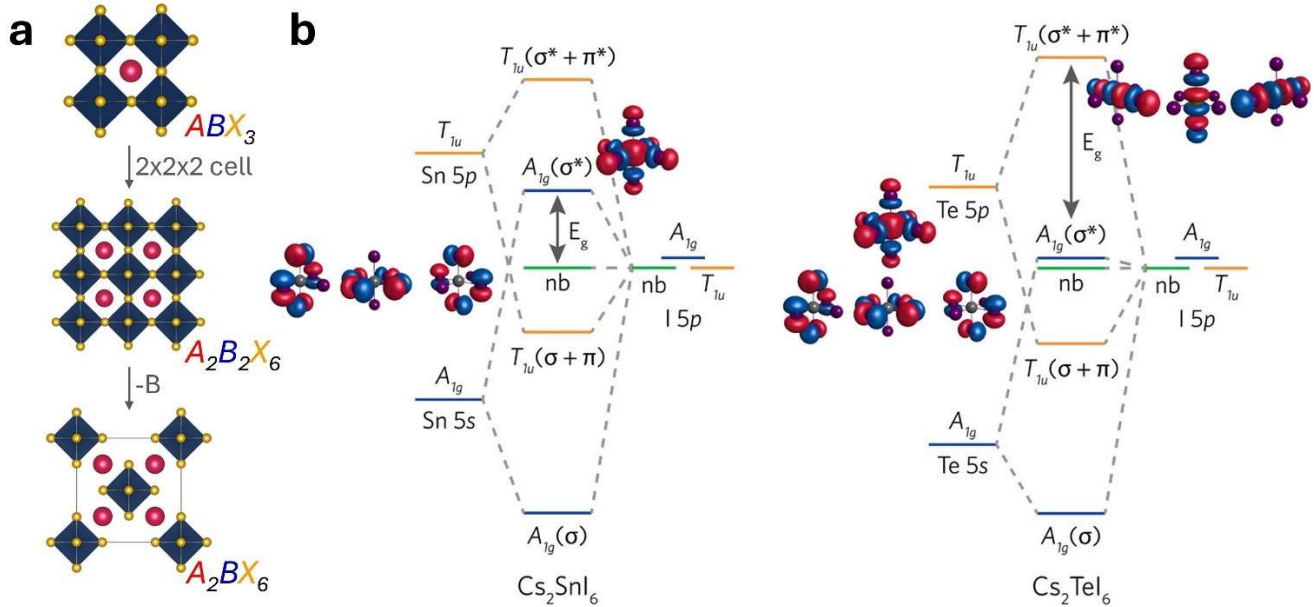

**Figure 3.** Structure and bonding of vacancy-ordered double perovskites. (a) relationship between the structures of conventional (ABX$_3$) and vacancy-ordered double (A$_2$BX$_6$) perovskites. (b) MO diagrams of Cs$_2$SnI$_6$ (left) and Cs$_2$TeI$_6$ (right). Molecular orbital plots were calculated on isolated SnI$_6$ and TeI$_6$ octahedra. Images adapted from ref. [74].

A key property of AMX$_3$ perovskites is the so-called defect tolerance, that is their ability to embody defects without generating trap states that promote fast non-radiative carrier recombination and/or quench the carrier mobility. Clearly, defect tolerance is highly beneficial both for light emission and light harvesting[77]. The question thus rises as to what features confer to ABX$_3$ perovskites this property. While multiple factors contribute[16,77], the antibonding nature of both VBM and CBM has been identified as a key feature, as discussed in the next section.

### 5. The importance of being antibonding.

Numerous semiconductors that can emit light efficiently and/or display good optoelectronic performance feature band edge states that consist of antibonding orbitals. This is the case for lead halide perovskites[16,66], including the so-called nD perovskites (n=0-2)[78,79], that are compounds featuring PbX$_6$ octahedra that extend in corner-sharing arrays in two dimensions (2D), one dimension (1D), or are structurally isolated from each other (0D). Interestingly, simple considerations on the MO diagram of PbX$_6$ can explain the antibonding nature of the CBM and VMB in all these compounds (Fig. 4). The general formula of simple n-dimensional perovskites is A$_{4-n}$PbX$_{6-n}$ (considering single rows or layers of corner-sharing octahedra for n=1 or 2, respectively). For 0D perovskites, since A are fully ionized monovalent cations (see Sect. 4), it is straightforward to see that each octahedron is PbX$_6^{4+}$, thus containing 10 (34) valence electron ignoring (considering) the X-p lone pairs perpendicular to Pb-X bonds. For such an electron configuration, Fig. 4 shows that the band edges are antibonding. As the octahedra become edge-sharing, the Pb-X molecular orbitals of adjacent octahedra combine into bands, i.e. they form closely spaced crystal orbitals whose bonding/antibonding nature depends on the k vector in the Brillouin zone[50]. However, the number of valence electrons remains equal to that of Pb(s)+X(s)+X(p) orbitals, thus the molecular orbitals that are occupied in an isolated PbX$_6^{4+}$ octahedron become filled bands in the nD perovskites, leaving the band edge states antibonding.

**Figure 4.** Schematic illustration of how the MO of a PbX$_6^{4-}$ octahedron becomes bands in APbX$_3$. On the left, the MO diagram of PbX$_6^{4-}$ as derived from textbook ligand field theory is shown. On the right, a pictorial DOS is represented with colored and empty semi-ellipses indicating occupied and empty states, respectively. X-p lone pairs perpendicular to the Pb-X bond are omitted in the MO diagram for clarity and shown in the pictorial DOS of APbX$_3$ in violet. Accordingly, π interactions between those X-p orbitals and Pb-p orbitals are not considered, as the partial electron density plots of ref. [66] show those interaction to be negligible (X-p orbitals perpendicular to Pb-X bonds behave mostly as lone pairs) and ref. [71] shows that the dispersion of the corresponding bands is about 20 times smaller than that of the Pb-X σ bonds, thus pointing towards localized electronic states typical of lone pairs.

Brandt et al. have shown that the antibonding character of the band edges of perovskites is a key property for defect tolerance[16,80]. In fact, a previous work showed that semiconductors with antibonding VBM and bonding CBM are likely to be defect tolerant[81]. That is, 'shallow' defects will predominantly form. These are defects whose electronic states lie close or even outside the band edges, rather than around the mid-band gap, where they would act as trap states and promote (non-radiative) charge recombination. The mechanism associated to the formation of these shallow defects can be understood from the schematic MO diagram of Fig. 5a. Considering that atom vacancies are the most common type of defect in semiconductors, and that these defects tend to create atomic-like energy levels in the neighboring atoms (dangling bonds), if the VBM/CBM are of antibonding/bonding nature, these defect level cannot fall far above/below the VBM/CBM, thus they will be shallow (Fig. 5a). For perovskites, however, the situation is different, as both VBM and CBM are antibonding. Brandt et al.[16] argue that the wide band dispersion in lead halide perovskites, promoted by the large spin-orbit coupling, makes the CBM falling below or close to the energy of Pb-6p atomic orbitals, thereby making the states associated to the dangling bonds shallow (Fig. 2b and 5b). Interestingly, this chemical bonding analysis was exploited to identify the electronic structure features that are likely to yield defect tolerant materials[16,80]. The Materials Project database[82] was screened to find materials fulfilling the identified criteria for defect tolerance, and the most promising materials were analyzed. Materials with long carrier lifetimes (> 1 ns) were successfully identified, thus demonstrating how a thorough, physically grounded chemical bonding analysis can lead to new strategies for materials design.

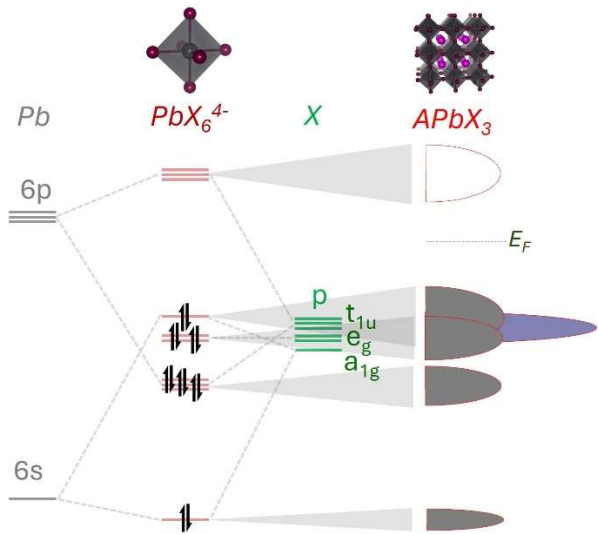

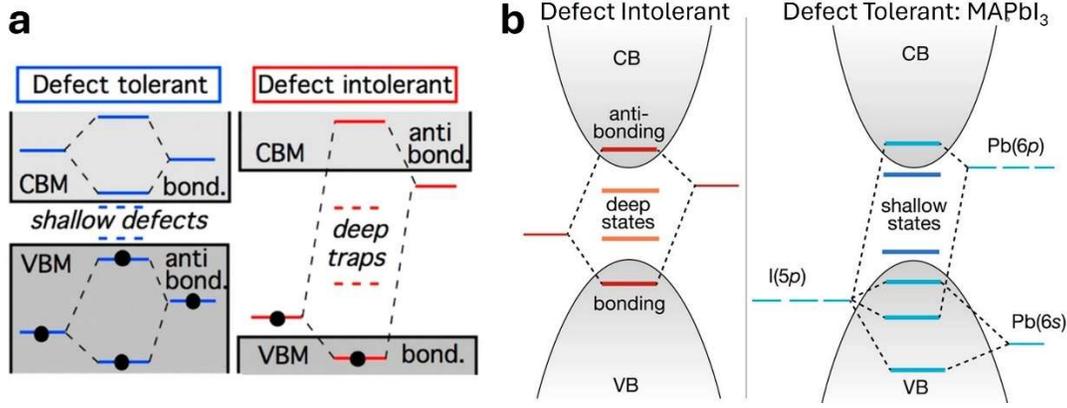

**Figure 5.** Schematic illustration of defect-tolerant and defect-intolerant chemical bonding patterns in the models of Zakutayev et al. for semiconductors in general [81] (a) and of Brandt et al. for perovskites [80] (b). Images taken from refs. [81] (a) and [80] (b).

In line with the findings discussed above, materials other than metal halides that exhibit good photovoltaic performance also display antibonding states at the band edges. In fact, this concept was exploited by Liu et al.[83] to fabricate a solar cell based on GeSe with notable power conversion efficiency (5.2%), which was maintained after prolonged air and irradiation exposure. This material was selected by virtue of its antibonding CBM/VBM, associated to good photovoltaic performance (see above). Moreover, they reasoned that the covalent bonding network of GeSe (anticipated from the small Ge-Se electronegativity difference) should overcome the poor environmental stability of lead halide perovskites, caused by the significant ionicity of their bonds[84]. This work is another example of how the knowledge derived from chemical bonding analysis can be leveraged towards materials design. Antibonding edge states are also found in $Cu_2ZnSnS_4$ (kesterite), a promising absorber material for solar cells. Indeed, Zhang et al.[85] demonstrated the antibonding nature of VBM ( S(3p)-Cu(3d) ) and CBM ( S(3p)-Sn(5p) ) through COHP, DOS and partial electron density analyses. Interestingly, by analyzing the chemical bonding in ordered and disordered $Cu_2ZnSnS_4$, they showed that interlayer Cu-Zn swapping, a commonly observed defect in this material, created electronic states close to the band edges (so-called 'band tails'). A comprehensive investigation based on non-adiabatic molecular dynamics simulations and electron-phonon coupling analysis showed how these band tail states are detrimental to the power conversion efficiency in photovoltaic devices. This understanding was then adopted to explain the mechanism through which Cd doping improves the solar cell performance, thereby opening the way for a rational design of kesterite-based materials for high-efficiency solar cells.[85]

The antibonding nature of VBM and CBM in materials appears to be highly beneficial also to promote light emission and high photoluminescence quantum yield (PLQY, i.e. the ratio between absorbed and emitted photons). Intriguingly, this is not (only) related to the defect tolerance mentioned above. Indeed, beside the high PLQY of lead halide perovskites, there are several instances of 0D metal halides in which substitutional doping makes the VBM and CBM antibonding and boosts their PLQY, for example: Te doping of $A_2SnCl_6$ (A=Rb, Cs)[76,86], Bi doping of $Cs_2SnCl_6$[87], and Sb doping in $Cs_2ZnCl_4$ [88]. As in all these cases the radiative recombination occurs at the doping centers, it appears implausible that the antibonding states improve the radiative recombination through defect tolerance, as defects are statistically more likely to form in the host lattice. It can be speculated that when both VBM and CBM are antibonding, they have a better overlap, which makes the electronic transition more likely (higher transition dipole moment), so that the radiative recombination is favored. However, time-dependent density functional theory simulations on Sb-doped $Cs_2ZnCl_4$[88] revealed that the electronic transition is allowed both in the host (VBM: Cl lone pairs, CBM: Zn-Cl antibonding) and in the guest (Sb-Cl antibonding in both VBM and CBM), despite the fact that the undoped material emits only weakly and only at low temperature. Thus, the role of antibonding orbitals in promoting radiative charge recombination cannot be explained by electronic transition probabilities only. Neither is it only a matter of defect tolerance (see above). It might involve dynamical effects not captured by static atomistic simulations, such as polarons formation or exciton trapping[89]. Certainly, this mechanism is not fully understood and deserves further investigation.

The discussion above highlighted how chemical bonding analysis, by identifying the correspondence between antibonding VBM/CBM and good materials performance (PLQY), can serve as a guide for the design of new light-emitting materials.

The antibonding character of the valence band was demonstrated to confer to semiconductors a low lattice thermal conductivity ($\kappa_L$)[90]. In particular, for a set of binary semiconductors, an inverse correlation was observed between the extent of antibonding character of the valence band, as estimated by COHP analysis, and the $\kappa_L$ of materials. This was due to the weaker nature of bonds when (p-d) antibonding orbitals are occupied, which makes these bonds softer. This results in a low speed of sound and high phonon-phonon scattering rates, two conditions that make $\kappa_L$ low. This finding was then exploited to develop screening criteria that led to the successful identification of 30 materials with low $\kappa_L$ from the database of experimentally determined crystal structures (ICSD[91]). While that work focused on thermoelectric materials, soft phonons and phonon-phonon scattering, beside leading to low $\kappa_L$, are highly beneficial for the photovoltaic performance of materials. Indeed, Yang et al.[92] showed that those phonon characteristics create a 'hot-phonon bottleneck' effect through which charge carriers are re-heated. This effect allows materials to overcome the intrinsic limit of power conversion efficiency (Shockley–Queisser[93]) in photovoltaic devices.

### 6. The role of $ns^2$ lone pairs.

Most semiconductors for optoelectronic applications contain heavy post-transition metals (In, Sn, Sb, Te, Tl, Pb, Bi), all of which share a peculiar feature: they can be stable as cations with electronic configuration $ns^2\ np^0$, beside the standard octet configuration ($ns^0\ np^0$). The presence of these $ns^2$ lone pairs in semiconductors has a wide range of structural and electronic implications that are linked to their optoelectronic properties and luminescence. Chemical bonding analysis can make that link intelligible.

A peculiarity of $ns^2$ lone pairs compared to the classical lone pairs encountered in organic chemistry textbooks is that they may or may not be "stereochemically expressed". That is, the atomic environment of a $ns^2$ cation can be symmetric (lone pair not expressed) or asymmetric (lone pair expressed)[15,94,95], as shown in Fig. 6a. Fabini et al.[70] argue that, even when $ns^2$ cations are in a symmetric environment, the 'unexpressed' lone pair manifests itself in the vibrational properties of the material. This is due to the lone pair producing a double-well shape of the potential energy surface of the compound. As proof of this concept, they point to the fluid-like Raman response and anharmonicity of lead halide perovskites and to the unusually high static dielectric constant of PbS. Note that these vibrational properties allow the formation of the polarons that play a key role in determining the outstanding optoelectronic performances of perovskites[96]. However, a subsequent computational and experimental study[97] comparing the electronic and vibrational properties of CsPbBr$_3$ and CsSrBr$_3$ concluded that, while the $ns^2$ lone pair is fundamental for many electronic and optical properties of perovskites, it is not required for the peculiar anharmonic vibrational behavior of perovskites. Rather, the vibrational properties are an intrinsic structural feature characteristic of halide perovskites.

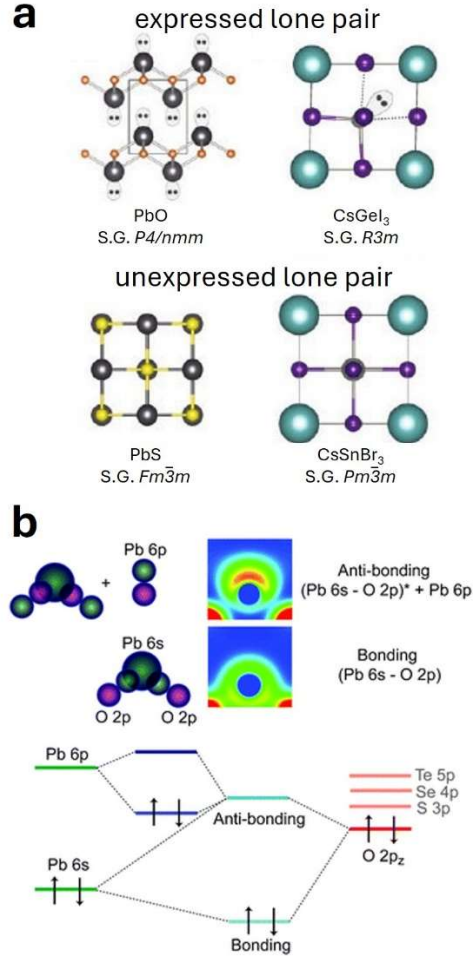

**Figure 6.** Structural and electronic origin of $ns^2$ lone pairs. (a) examples of structures where the $ns^2$ lone pair is (un)expressed. (b) Orbital interaction mechanisms leading to the appearance of a 'stereochemically expressed' $ns^2$ lone pair in PbO. Images taken from refs. [70] (a; reproduced with permission of Springer Nature) and [98] (b; used with permission of the Royal Society of Chemistry; permission conveyed through Copyright Clearance Center, Inc.).

Walsh et al.[98] put forward an intriguing model to explain the stereochemical expression of $ns^2$ lone pairs through chemical bonding arguments. This model is illustrated in Fig. 6b on PbO, a prototypical example. The occupied antibonding orbital resulting from the interaction between Pb(s) and O(p) orbitals can hybridize with the empty Pb(p) orbital. This is only possible when the Pb does not lie in a high-symmetry position, *i.e.* when the O-Pb-O angle deviated from 180°. This hybridization clearly stabilizes the orbital and lowers the overall energy of the system. However, for it to occur, the compound has to undergo a distortion, which tends to be energetically unfavourable, typically due to the reduced coordination of the cation. Whether the $ns^2$ lone pair is 'stereochemically expressed' is thus determined by i) the relative energies of cation s and anion p states, which in turn determine how strongly they hybridize, and ii) the competition between electronic stabilization of the antibonding orbital and the energy penalty associated to the deviation from the high symmetry coordination. This scenario is expected to have general validity and thus to apply to halide perovskites as well. Interestingly, this explanation rationalizes a key structural feature of optoelectronic materials in terms of chemical bonding arguments, in particular orbital hybridization. Ogawa et al.[99] demonstrated the predictive power of this mechanism by studying the structural consequences of $Y^{3+}$ to $Bi^{3+}$ substitution (thus introducing a $ns^2$ lone pair cation with a similar cation radius) in $Bi_2YO_4X$ (X=Cl, Br, I). They observed a band gap reduction and other changes in the electronic and vibrational properties that are in agreement with the above mentioned model.

In many semiconductors adopted for optoelectronic applications, the presence of cations with a $ns^2$ lone pair leads to an overall electron configuration in which the band edge states are antibonding. This can be clearly seen from Fig. 4 and the related discussion, and also in Fig. 3, where the substitution of a $Sn^{4+}$ ($s^0p^0$ configuration) ion with a $Te^{4+}$ ($s^2p^0$ configuration) ion changes the band edge states from lone pairs and s-p antibonding to s-p and p-p antibonding. The question thus arises as to whether the excellent light emission and harvesting properties discussed in Sects. 4 and 5 for perovskites and other metal halides are the result of antibonding orbitals or of the presence of $ns^2$ cations. To the best of our knowledge, this fundamental question has never been addressed. We speculate that these two features may act in synergy, for example the lone pair could act on the vibrational properties that are beneficial for carrier lifetime and mobility (e.g. by promoting polarons formation[100]), while the antibonding states could, beside conferring defect tolerance[16,80], affect the dynamics of carrier recombination. Insights on this topic can be of fundamental importance for the discovery of new toxic-elements-free semiconductors with the same outstanding optoelectronic properties of perovskites.

Finally, we mention that the presence of cations with $ns^2$ lone pairs is not necessarily beneficial for the optoelectronic performance, as demonstrated by Kim et al. in the kesterite $Cu_2ZnSnS_4$[101]. There, Sn is in a +4 oxidation state, thus with a $s^0p^0$ configuration. Through systematic simulations of defects formation energies and ionization levels, it was shown that the ease with which $Sn^{4+}$ is reduced to $Sn^{2+}$ leads to the formation of defects with deep energy levels and large carrier capture radius, which promotes charge recombination and kills the energy conversion process. The authors proposed that similar mechanisms may explain the poor photovoltaic performance of other materials containing lone pair cations.

Given their importance, $ns^2$ lone pairs are widely studied in the literature of optoelectronic materials. We focused on those studies that discuss this phenomenon from the chemical bonding perspective. For more comprehensive reviews, we refer the reader to refs. [94,95,102].

### 7. Metal Chalcohalides: interplay of different chemical bonds

Nearly all materials discussed above have their (opto)electronic properties determined by only one type of chemical bond (e.g. the M-X bond in metal halides $A_aM_mX_x$, with the A cation serving a primarily structural role). The interplay of several types of chemical bonds in a single material can give rise to properties and phenomena absent in single-bond systems. Chalcohalides, being composed of metal cations and both chalcogen and halogen anions, represent an example of this interplay that bears relevance for optoelectronic applications. This family of materials is being intensively studied as sustainable alternative to lead halide perovskites. Indeed, some metal chalcohalides exhibit electronic features similar to lead halide perovskites, such as low carrier masses, defect tolerance, and IR-visible band gaps[103]. Their key advantage over lead halides, beyond the possibility of being made of low-toxicity and earth-abundant elements, is their superior environmental stability. This finds its roots in the chemical bonding pattern: given the smaller cation-anion electronegativity difference, the chemical bonding in chalcohalides is more covalent than in lead halides, thus making the former less prone to decomposition and moisture-induced degradation[103,104]. In this section, we discuss reported studies that focus on the chemical bonding in this emerging class of materials.

The mixed anion nature of chalcohalides generally produces an anisotropic environment for cations, which often results in dimensionality reduction of the overall crystal structure. A typical example is the isostructural series of PnChX chalcohalides (Pn=Sb, Bi ; Ch=S, Se ; X= Cl, Br, I), that is formed by covalently bonded $[Pn_2Ch_2X_2]_n$ chains held together by van der Waals (vdW) interactions [105,106] (Fig. 7). This anisotropy is also manifested in the needle-like shape of the corresponding crystals[106,107]. Caño et al. suggested that the formation of low-dimensional compounds can be predicted based on the (Pauling) electronegativity difference between cations and anions, with vdW compounds forming when this difference is lower than 1.5 [105,108]. Chemical bonds anisotropy impacts vibrational properties as well. In fact, in $CuBiSCl_2$, the mixed anion environment of Cu atoms is directly responsible for the low lattice thermal conductivity of this material[109]. There, Cu atoms have an elongated octahedral coordination with two relatively strong, axial Cu-S bonds and four weak equatorial Cu-Cl bonds. Consequently, in the equatorial plane, Cu atoms exhibit wide, possibly rattling motion, responsible for the low $\kappa_L$ of the material. This mechanism is analogous to that discussed in Sect. 5[90], as occupied Cu-Cl antibonding states are found right below the Fermi level. We note that in this study and in a similar work on $Pb_mBi_2S_{3+m}$ chalcohalides[110], the 1D structural character is attributed to the $ns^2$ lone pair expression rather than to the electronegativity difference. As the dimensionality reduction is determined by the chemical bonds of the material and has a direct impact on technologically relevant properties, chemical bonding investigations specifically aimed at explaining this phenomenon will be highly beneficial for the rational design of chalcohalide materials.

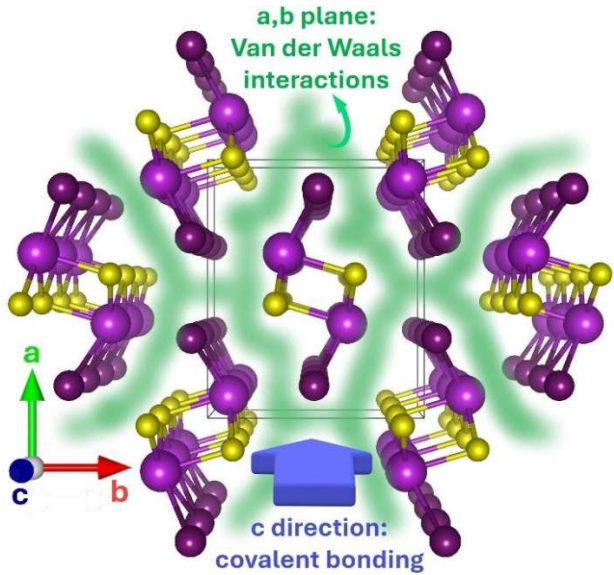

**Figure 7.** Crystal structure of BiSI, highlighting its 1D structure. Van der Waals interactions are pictorially indicated as a green halo, while the blue arrow indicates the direction along which atoms are covalently bonded. Bi, S, and I atoms are colored in light violet, yellow, and dark violet, respectively. Structure drawn with VESTA software[111].

Crystallographically different Bi atoms form different types of chemical bonds, some of them appearing as $Bi_2$ dimers (Fig. 8a). Electronic structure and chemical bonding analysis demonstrated[13] that this material contains Bi atoms in two different oxidation states, +3 and +2, the latter forming dimers. These dimers are thus *hitherto* unobserved $Bi_2^{4+}$ chemical entities. They give rise to a mid-gap bonding state (Fig. 8b), responsible for a weak infrared absorption peak observed in this material. The fascinating aspect is that these $Bi^{2+}$ atoms behave as a textbook case of Peierls distortion[112,113], namely as a chain of atoms, each with one valence electron, that becomes stabilized by dimerization. This is supported by i) p-DOS, showing that Bi(s) states lie 8 to 15 eV below the VBM, thus they are unlikely to participate into chemical bonds, leaving $Bi^{2+}$ cations with one valence electron in a 6p orbital, and ii) a model simulation with $Bi^{2+}$ atoms equidistant from each other, which, beside being unstable, gives rise to a metallic state, just like in the Peierls model. We note that a recent study put forward the same Peierls distortion scenario in the structurally analogous $Bi_{13}S_{18}I_2$ compound[114]. This result has also important implications for the behavior of elements with $ns^2$ lone pairs, as in this material they behave as if they are part of the core orbitals. Therefore, the behavior of $ns^2$ lone pairs is strongly dependent on the chemical environment of the cation, along the same line of the model shown in Fig. 6.

$Bi_{13}S_{18}Br_2$ represents an interesting case of chemical bonding interplay which gives rise to peculiar phenomena[13].

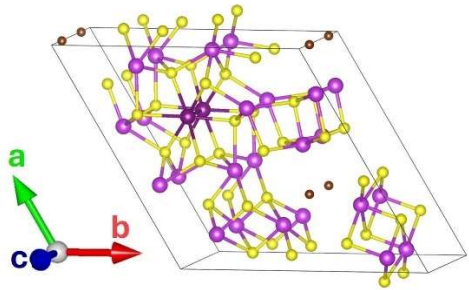
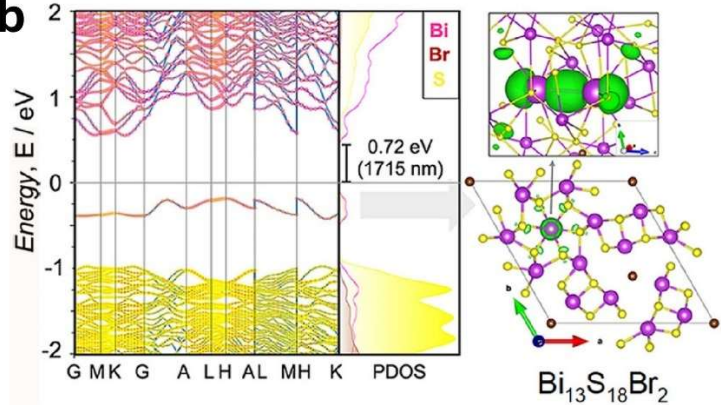

**Figure 8.** Bi dimers in $Bi_{13}S_{18}Br_2$. (a) crystal structure of $Bi_{13}S_{18}Br_2$; Yellow, brown, and violet spheres represent, respectively, S, Br, and Bi atoms. Bi atoms forming the dimer are colored in dark violet. (b) band structure of $Bi_{13}S_{18}Br_2$, with bands colored according to the atom contributions. The electron density relative to the valence band is shown on the right as green isosurfaces, in two different orientations. Panel b adapted from ref. [13].

Antimony sulfoiodide $Sn_2SbS_2I_3$ has been the subject of several studies due to both its good photovoltaic performance and its elusive properties[115–117]. In particular, detailed computational and crystallographic analyses revealed the crystal structure of this compound to be an average over multiple polar configurations and to display both static and dynamic disorder[116,117]. Another puzzling aspect is the anomalous increase in band gap in passing from $Pb_2SbS_2I_3$ to $Sn_2SbS_2I_3$, which however can be explained through orbital hybridization arguments. The relativistic contraction of Pb(s) orbitals makes them lower in energy than Sn(s) orbitals, thus the latter can better hybridize with the anion p orbitals, which lie at higher energy (see *e.g.* Figs. 2 and 4)[116]. This results in the cation(s)-anion(p) antibonding states forming the valence band moving higher in energy for Sn compared to Pb, thereby reducing the band gap. In this material, both band edges are formed by antibonding orbitals, similarly to perovskites. Consequently, in line with the discussion on defect tolerance of Sect. 5, it was observed that the intrinsic disorder of the material does not produce deep trap states[117]. We mention that also $Sn_2SbS_2I_3$ crystallizes in the 1D structure[116] typical of chalcohalides such as BiSI (Fig. 7).

$Ag_3XY$ (X = S, Se; Y = Br, I) compounds display a band gap reduction as T increases, contrary to most materials[118]. To explain this

phenomenon, Benítez et al.[119] resorted to molecular dynamics simulations and tight-binding Hamiltonians (TBH). In a nutshell, TBH are effective models in which the electronic structure and/or energy of the material are expressed in terms of localized atomic orbitals, with parameters describing the on-site energies and hopping interactions (we refer the reader to ref.[120] for a pedagogical introduction to TBH and ref.[121] for the Wannier-based type of parametrization adopted in the work on $Ag_3XY$). Through this approach, it was shown that phonon-induced atomic displacements enhance the Ag(s)-S(s) interaction, thereby widening their bonding-antibonding gap. As the corresponding bonding states form the conduction band of the material, their energy lowering reduces the band gap. An important aspect of this study is the explanation of material properties through TBH. As TBH define a Hamiltonian representation of the system, they are conceptually different from the bond characterization methods discussed in Sects. 2 and 3. Nonetheless, it is fascinating that some quintessential chemical bonding concepts such as bond strength, electron delocalization, and even lone pairs[122] appear also in TBH. Works that aim at combining TBH and chemical bonding analysis into a unified methodological framework are appearing in the literature, see for example [123].

Chalcohalides have been known for over six decades, but their recently uncovered optoelectronic properties have renewed interest in these materials[103]. Their chemical bonding is complex and only relatively few studies analyzed it. In the future, a more systematic investigation can both uncover new intriguing chemical phenomena (such as in the $Bi_2^{4+}$ dimers[13]) and guide the rational design of sustainable chalcohalide materials for applications such as light harvesting and emission.

### 8. The role of chemical bonding analysis in the future of materials science.

The previous four sections showcased how the knowledge brought about by chemical bonding analyses can produce a form of *predictive understanding* on the optoelectronic properties of materials. Let us elaborate on this expression. Rigorous, prediction-oriented chemical bonding analysis can be leveraged to build models that connect the composition and/or the structure of a material to its properties. On the one hand, these models explain *why and how* a given characteristic of a material produces a given property, thereby creating a general form of *understanding*. For example, it was shown how the defect tolerance of materials can be traced back to their chemical bonding pattern (Sect. 5), or why many materials with good optoelectronic performance contain heavy post-transition metals (Sect. 6). We also illustrated, in fairly complex materials, the impressive explanatory and predictive power of MO diagrams, a tool originally devised for molecules. On the other hand, these models can *predict*, at least on a qualitative level, the properties of unknown materials. That is, the models resulting from chemical bonding analysis can afford a form of materials design, or the initial stage thereof. Successful examples of materials design protocols directly built on chemical bonding analysis have been discussed in this perspective, such as the works on high carrier lifetime and on low thermal conductivity of Sect. 6.

In line with the reasoning above, we argue that a wider adoption of prediction-oriented chemical bonding analysis can advance materials science and strengthen its foundations. The tools of modern computational chemistry offer immense potential to build predictive, quantum-mechanically-rooted chemical bonding models and rules specifically designed for solids. In particular, we refer to the large computational power, which makes accurate solid-state electronic structure simulations nearly a routine task, and to the direct availability of a vast amount of materials data[124], a precious resource to derive models and appraise their general validity. In fact, predictive models for solids have been devised since the early days of theoretical chemistry, for example the Hume-Rothery rules[125] and Pauling's rules[126]. However, the aforementioned tools can be exploited to expand that legacy towards more accurate and physically rooted chemical bonding models. For example, a recent study[127] analyzing a materials database showed that, out of ~5000 experimentally known oxides, only 13% of them fulfilled all Pauling rules. This example showcases how the tools of present-day computational chemistry (the availability of big data in this case) can indicate when historical models need to be rethought. In our view, new solid-state chemical bonding models should be systematically developed, which: i) build on the existing chemical bonding knowledge, improving, expanding, or even abandoning existing models when necessary, ii) have a close connection to the quantum mechanical principles of the electronic structure, and, iii) have predictive power. Requirements ii) and iii) can, in some cases, require a tradeoff between predictive accuracy and fidelity (see Sect. 2 and below). In these regards, it should be borne in mind that chemistry occupies a special epistemological position in science, for its principles being at the same time heuristic and based on the fundamental laws of physics. This duality is widely discussed in the philosophy of science[25,128–132]. In fact, the modern availability of accurate electronic structure information for a large number of materials could be exploited to build chemical bonding models that embody the quantum mechanical mechanisms underpinning the formation of chemical bonds. Nonetheless, we believe that the guiding principles of chemical bonding models should be their predictive ability, especially when one needs to discriminate among conflicting models. A representative example of a chemical bonding model that fulfills the requirements above is the lone pair model discussed in Sect. 6 (Fig. 6): it explains the properties of materials, it is predictive, and it is also linked to the quantum mechanical mechanism of bonding. Encouraging studies in the direction of prediction-oriented chemical bonding models for materials, besides those discussed in this perspective on optoelectronic materials, have recently appeared in the literature, for example on the relationship between chemical bonding and certain material properties (*e.g.* electrical conductivity, band gap, melting point)[133] and on the usage of chemical bonding analysis in the design of thermoelectric materials[134].

Finally, a central question arises: given the increasing capabilities of machine learning (ML) algorithms in predicting the properties of compounds, what role should chemical bonding analysis play in the future of chemistry and materials science? The topic of human vs artificial intelligence in computational chemistry has been largely discussed in the literature, and we refer the reader to the interesting opinions of Hoffmann and Malrieu[135–137], and George and Hautier[138].

We address the opening question with a threefold answer, avoiding replicating those previous discussions. The first answer is implicit in the paragraphs above: chemical bonding analysis brings an understanding on the behavior of chemical species, thereby creating what is generally defined as chemical knowledge. This knowledge constitutes the fabric of chemistry, making it a teachable subject and a field of scientific inquiry. The synthetic procedures might in the future be largely performed by the AI-driven robots of self-driving labs[139]. We briefly mention that explainable and interpretable ML models exist in materials science[140], for example the discovery of chemical laws through symbolic regression[141,142]. While increasingly valuable in fostering our understanding of material properties, at present they do not bring the kind of understanding (see definition above) that forms the backbone of chemistry. A second answer is that, even ignoring the knowledge-creating role chemical bonding analysis, its predictive power cannot be fully replaced by ML alone. Indeed, ML has important limitations. ML models struggle to extrapolate beyond the data on which they were trained, limiting their predictive power to known chemical spaces[143]. Creating ML models with general validity is in principle possible, but the amount of (accurate) data required grows substantially, making it conceivable that such large amounts of reliable data may never become available[144]. The third and arguably most relevant answer is that the role of chemical knowledge can be to form a synergy with ML algorithms and artificial intelligence in general. This synergy can take many forms. The most straightforward one is to embed chemical bonding concepts into the feature representation for ML, as already customarily done for example when predicting the properties of compounds based on their chemical formula[138,145–147]. A key challenge in this regard is to find the most effective way to frame chemical bonding models when embedded into ML algorithms[132]. We note that more ambitious, large-scale efforts to combine chemical bonding and ML in the prediction of material properties are being planned[148]. Another type of synergy occurs when using ML as a tool to improve the accuracy of quantum chemical simulations. ML force fields[149] for molecular dynamics are a typical example: ML would be used to improve the accuracy and/or reduce the computational cost of the simulations, but these would still be analyzed in the a chemical framework to reveal the mechanism of the reactions and other atomistic insights. Overall, this discussion on ML aimed at highlighting possible interplays with chemical bonding analysis, without expecting to be exhaustive, as ML in chemistry and materials science is a vast and rapidly evolving field. Game-changing ML developments, with possible implications for theoretical chemistry, might easily lie ahead.

Despite being over a century old, chemical bonding remains a concept with transformative yet underexplored potential for advancing materials science. Chemical bonding analysis can foster the development of materials chemistry while making it more engaging for the chemistry community at large, as illustrated in this work for optoelectronic materials. We trust that this perspective will spark a broader adoption of this fascinating foundational concept of chemistry.

## AUTHOR INFORMATION

### Corresponding Author

Gabriele Saleh - Nanochemistry Department, Istituto Italiano di Tecnologia, Via Morego 30, 16163 Genova, Italy; gabriele.saleh@iit.it
Liberato Manna - Nanochemistry Department, Istituto Italiano di Tecnologia, Via Morego 30, 16163 Genova, Italy; liberato.manna@iit.it

### Author Contributions

The manuscript was written through contributions of all authors. All authors have given approval to the final version of the manuscript.

## ACKNOWLEDGMENT

The authors acknowledge funding from the Project IEMAP (Italian Energy Materials Acceleration Platform) within the Italian Research Program ENEA-MASE (Ministero dell'Ambiente e della Sicurezza Energetica) 2021–2024 "Mission Innovation" (ageement 21A033302 GU No. 133/5-6-2021) and from the European Research Council through the ERC Advanced Grant NEHA (grant agreement n. 101095974).

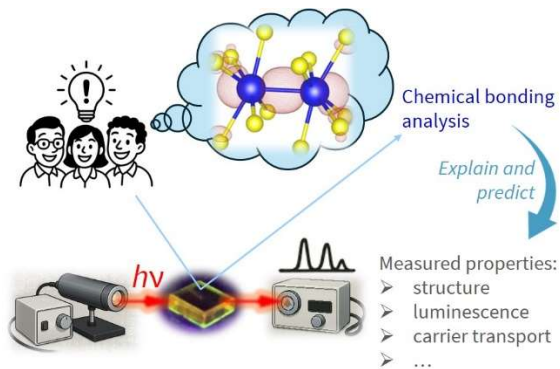

---

* $Cs_2TlBiBr_6$ perovskite does have a band structure similar to $CsPbBr_3$. However, there are no report of $Cs_2TlBiBr_6$ being synthesized in the perovskite structure. Besides, Tl is extremely toxic (more than lead).